\documentclass[a4paper]{article}

\usepackage{INTERSPEECH2020}
\usepackage{multirow}
\usepackage{verbatim}
\usepackage{graphicx}
\usepackage{array}
\title{The HCCL Speaker Verification System for Far-Field Speaker Verification Challenge }
\name{Zhuo Li$^{1,2}$, Ce Fang$^{1,2}$, Runqiu Xiao$^{1,2}$, Zhigao Chen$^{1,2}$, Wenchao Wang$^{1,2}$, Yonghong Yan$^{1,2,3}$}
\address{
 $^1$Key Laboratory of Speech Acoustics and Content Understanding, Institute of Acoustics, Chinese Academy of Sciences, Beijing, China\\
 $^2$University of Chinese Academy of Sciences, Beijing, China\\
 $^3$Xinjiang Key Laboratory of Minority Speech and Language Information Processing, Xinjiang Technical Institute of Physics and Chemistry, Chinese Academy of Sciences, Urumqi, China
}
\email{\{lizhuo,fangce,xiaorunqiu,chenzhigao,wangwenchao,yanyonghong\}@hccl.ioa.ac.cn}
\begin{document}

\maketitle

\begin{abstract}
This paper describes the systems submitted by team HCCL to the Far-Field Speaker Verification Challenge. Our previous work in the AIshell Speaker Verification Challenge 2019 shows that the powerful modeling abilities of Neural Network architectures can provide exceptional performence for this kind of task. Therefore,  in this challenge, we focus on constructing deep Neural Network architectures based on TDNN, Resnet and Res2net blocks. Most of the developed systems consist of Neural Network embeddings are applied with PLDA backend. Firstly, the speed perturbation method is applied to augment data and significant performance improvements are achieved. Then, we explore the use of AMsoftmax loss function and propose to join a CE-loss branch when we train model using AMsoftmax loss. In addition, the impact of score normalization on perfomance is also investigated. The final system, a fusion of four systems, achieves minDCF 0.5342, EER 5.05\% on task1 eval set, and achieves minDCF 0.5193, EER 5.47\% on task3 eval set.

\end{abstract}

\section{Introduction}
In the past decade, with the development of deep learning, speaker verification technology has significantly advanced in telephone channel and close-talking. Recently, due to its extensive applications in smart home and smart city, speaker verification under far-filed and complex environment has aroused great interest in research and authentic applications.  However, with the adverse impacts of the long-range fading, room reverberation, and complex environmental noises, etc., far-field speaker verification is still challenging.

In this paper, we describe the speaker verification systems developed by the team HCCL for the FFSVC2020 challenge\cite{qin2020ffsvc}. The challenge aims to benchmark the state-of-the-art speaker verification technology under far-field and noisy conditions, meanwhile, promote the development of new ideas and techniques in speaker verification. The challenge has three tasks, far-field text-dependent speaker verification from single microphone array, far-field text-independent speaker verification from single microphone array and far-field text-dependent speaker verification from distributed microphone array. Our team mainly participate in task1 and task3. 

According to our prior work in the AIshell Speaker Verification Challenge 2019\cite{qin2019hi}, we explore several deep speaker embedding extractors based on TDNN(Time Delay Neural Network)\cite{snyder2015time}, Resnet(Residual Neural Network)\cite{he2016deep} blocks and Res2net\cite{gao2019res2net} blocks in this challenge. In addition, the impact of data augmentation, loss function, backend strategies and score normalization techniques on systems performance is also analyzed.

\section{System description}
This section describes the systems we develope for the FFSVC2020 challenge. Firstly, we introduce the datasets, data augmentation and spectral feature used in model training. Secondly, we introduce several state-of-the-art models, including the common TDNN xvector architecture\cite{snyder2017deep}, ETDNN(Extended TDNN) xvector architecture\cite{Daniel2018etdnn}\cite{Daniel2018etdnn2}, FTDNN(Factorized TDNN) xvector architecture\cite{Daniel2018Semi}, Resnet\cite{he2016deep}\cite{cai2018exploring}\cite{villalba2020state}\cite{gusev2020deep} architecture and Res2net\cite{he2016deep} architecture. Specially, we explore a unique resnet model, Res2net50 model, which proposed in paper\cite{gao2019res2net}. Thirdly, we explore the use of AMsoftmax(the Additive Margin Softmax)\cite{amsloss2018} loss to improve the performance. Finally, some backend methods, such as PLDA(Probabilistic Linear Discriminant Analysis)\cite{ioffe2006probabilistic}\cite{prince2007probabilistic}, EDA(Enrollment Data Augmentation)\cite{qin2019hi}, ASnorm(Adaptive Symmetric Score Normalization)\cite{cumani2011comparison} and scores fusions are described.

\subsection{Training data preparation}

The training sets used in our experiments are VoxCeleb-1+2\cite{nagrani2017voxceleb}\cite{chung2018voxceleb2}, AIshell\cite{bu2017aishell}\cite{du2018aishell-2}, HIMIA\cite{qin2019hi} and part of the DMASH Dataset that FFSVC2020\cite{FFSVCbaseline} provides. 

The challenge provides a training set with 120 speakers, and a development set with 35 speakers. The HIMIA dataset has 254 speakers in the training set and 42 speakers in the developement set. There are 405 speakers in total after excluding duplicates in two data sets. For each text-dependent utterance in the HIMIA dataset, we use the recordings from four recording devices, which include one 25cm distance microphone and three randomly selected microphone arrays (4 channels per array). For the sake of clarity, the datasets notations are defined as in table~\ref{dataset}.

In order to increase the number of speakers in the HIMIA dataset and FFSVC dataset, the speed perturbation\cite{ko2015audio}\cite{yamamoto2019speaker} method is introduced to creat copies of the original signal with speed factors of 0.8, 0.9 and 1.1 using standard Kaldi\cite{povey2011kaldi} speed perturbation recipe. Because the speed perturbation results in changed speech spectrum, the copies are considered to belong to another speaker which differs from the original. In this way, the number of speakers grows a lot.

For the purpose of increasing the amount and the diversity of the training data, all training data, including copies created by the speed perturbation, is augmented by using the freely available MUSAN\cite{snyder2015musan} and RIRs datasets, creating four corrupted copies of the original recordings with Kaldi recipe.

\begin{table}[hpbt]
\vspace{-2mm}
\centering
\caption{Datasets Notations}\label{dataset}
\vspace{-5mm}
\begin{center}
\small
\setlength{\tabcolsep}{1.3mm}{
\centering
\begin{tabular} {|c|p{50pt}|p{60pt}|c|}\hline
notation & datasets & \multicolumn{1}{c|}{augmentation} & num spks \\\hline
Vox & Voxceleb 1+2 & \multicolumn{1}{c|} {noise aug} & 7363 \\\hline
Vox1 & Voxceleb 1 & \multicolumn{1}{c|}{noise aug} & 1211 \\\hline
AIshell & AIshell & \multicolumn{1}{c|}{noise aug} & 2401 \\\hline
FFSVC405 & FFSVCdata + HiMIA & \multicolumn{1}{c|}{noise aug} & 405 \\\hline
FFSVC1215 & FFSVCdata + HiMIA & noise + speed 0.9, 1.1 aug & 405*3 \\\hline
FFSVC1620 & FFSVCdata + HiMIA & noise + speed 0.8, 0.9, 1.1 aug & 405*4 \\\hline
\end{tabular}}
\end{center}
\vspace{-8mm}
\end{table}

\subsection{Feature extraction}

All training datasets are resampled to 16kHz and pre-emphasized before feature extraction. 30-dimensional MFCCs (Mel Frequency Cepstral Coefficients) extracted from 25ms frames with 10 ms overlop, spanning the frequency range 20Hz-7600Hz and 64-dimensional MFB (Log Mel-filter Bank Energies) from 25ms frames with 10ms overlop, with frequency limits 0-8000Hz are used in this challenge.

\subsection{Embedding Extractors}

The systems based on TDNN used by our team include xvector, ETDNN xvector, FTDNN xvector. The systems based on Resnet blocks use in this challenge name Resnet34 and Resnet50.

\begin{figure}[!htbp]
  \centering
  \includegraphics[width=75mm]{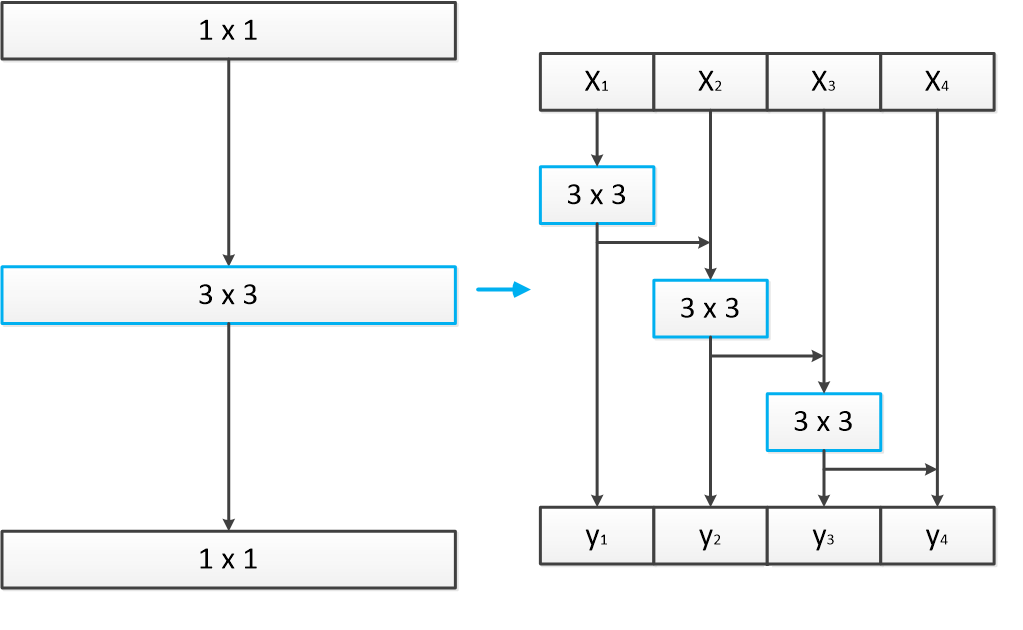}
  \vspace{-2mm}
  \caption{Simplified Res2Net Block}
  \label{res2net_fc}
  \vspace{-6mm}
\end{figure}

Res2net block is a modified version of Resnet block, as shown in Figure~\ref{res2net_fc}. The 3*3 filters are replaced by several smaller filter groups, and these outputs are concatenated after convolution operations. There are two important parameters, \textit{width} and \textit{scale}. \textit{Scale} represents the number of groups that one $3*3$ filter is split into. \textit{Width} represents the number of channels of each small group. For details, please refer to the paper \textit{A unique multi-scale architecture for text-independent speaker verification}, which is submitted by our laboratory to interspeech2020.

\subsection{Loss Function}
Recent studies have shown that AMsoftmax loss has greatly improved performance in the field of speaker verification\cite{liu2019large}. We also use AMsoftmax in this challenge. Meanwhile, we add CE-loss branch to the network to do joint training. The AMsoftmax loss function is formulated as:
$$L_{AMS} = \frac{1}{N}\sum_{i}-log\frac{e^{s*(cos\theta_{y_i} -m)}}{e^{s*(cos\theta_{y_i} -m)} + \sum_{j\not=y_i}e^{s*cos(\theta_{y_j},i)}}$$
where $s$ is a scale factor and $m$ is the margin factor.
\subsection{Backend}
In this work, either CS(cosine similarity) or PLDA is used for scoring. Additionally, EDA and ASnorm are also applied.
\subsubsection{Enrollment Data Augmentation}
There is a large mismatch between the enrollmenat and the test utterances because of the different recording environments. Recent researches have shown that it's an effective method to relif this mismatch by using MUSAN and RIRs dataset to augment enrollment utterances\cite{qin2019hi}\cite{FFSVCbaseline}. The final enrollment embedding is obtained by averaging the embedding from the original uttrance and the augmented versions.
\subsubsection{Score Normalization}
Finally, ASnorm scheme is also adopted as proposed in \cite{cumani2011comparison}. For every pair $(x_1,x_2)$, the normalized score can be calculated as follows:
$$\hat{S}(x_1,x_2) = \frac{S(x_1,x_2) - \mu_1}{\sigma_1} + \frac{S(x_1,x_2) - \mu_2}{\sigma_2} $$
Here, $\mu_1$  and $\sigma_1$ are calculated by matching $x_1$ against the cohort set and similarly for $\mu_2$ and $\sigma_2$. Means and standard deviations are calculated using a set of $N$ best scoring impostors.

\section{Implementation, Result and Analyze}

We develop three series of systems based on TDNN and Resnet blocks, including Resnet systems used pyTorch, TDNN systems used Kaldi\cite{povey2011kaldi}, TDNN systems used pyTorch. The first series are built using pyTorch, and based on Resnet and Res2net model. The second series are built using the Kaldi toolkit, and the TDNN and ETDNN xvector models are first pre-trained, then FFSVC2020 train-dataset and HIMIA dataset are used to finetune the model. 
The third series are built using pyTorch, and the ETDNN and FTDNN xvector models are trained only using FFSVC2020 train-dataset and HIMIA dataset. 
In addition, we explain the naming rules which is followed throughout this paper for all systems. The naming rules are divided into two categories, finetune and non-finetune. The first is named as $\langle$\textit{the training data used for base model}$\rangle$-$\langle$\textit{feature}$\rangle$-$\langle$\textit{architecture}$\rangle$-$\langle$\textit{the training data used for finetune}$\rangle$-$\langle$\textit{feature}$\rangle$-\textit{finetune}, and the second is named as $\langle$\textit{the training data used for model}$\rangle$-$\langle$\textit{feature}$\rangle$-$\langle$\textit{architecture}$\rangle$(-$\langle$\textit{loss function}$\rangle$).

\subsection{Resnet systems used pyTorch}
All systems in this series are implemented with pyTorch and use MFB as input feature. Meanwhile, dropout is applied to embedding layer, and CS is used for scores in this series. In addition, EDA is used in all results, including other series.

\textbf{Vox1-MFB-Resnet34-FFSVC405-MFB-finetune:} Considering the number of speakers in FFSVC405, finetune is a good transfer learning method for this challenge. The Resnet34 model, which described in \cite{cai2018exploring}, is pre-trained with 1211 speakers in Vox1-MFB.
The widths (number of channels) of the residual blocks used in our experiments are \{32,64,128,256\}. In this stage, the model is trained using stochastic gradient descent with weight decay 1e-4 and momentum 0.9. The learning rate is set to 0.1, 0.01, 0.001 and is switched when the training loss plateaus. After the pre-trained model converges, FFSVC405-MFB is used to finetune the model with the learning rate 0.001. 

\textbf{FFSVC405-MFB-Resnet34:} For comparison, the resnet34 model is directly trained with FFSVC405-MFB. Stochastic gradient descent optimization is used with weight decay 1e-4 and momentum 0.9. The learning rate is set to 0.1, 0.01, 0.001 and is switched when the training loss plateaus.

In order to facilitate the training of PLDA, the embedding-size is expanded from the original 128 dimensions to 512 dimensions. The impact of embedding size on performance is also explored by comparative experiments. Meanwhile, Spectrum masking\cite{park2019specaugment} is used in our training stage, and masking 5\%-10\% of the Spectrogram for each utterance. The results are summarized in Table~\ref{R1}, and all scores in this Table use CS for scoring. It could find that embedding size has no effect on system performance, spectrum masking can yield a 3\%-5\% relative improvement in EER and minDCF, finetune can yield about 8\% relative improvement in EER and 2\%-4\% improvement in minDCF. 
\begin{table}[!hbt]
\vspace{-2mm}
\centering
\caption{Research of mask and embedding size}\label{R1}
\vspace{-5mm}
\begin{center}
\small
\setlength{\tabcolsep}{1.3mm}{
\begin{tabular}{c|c|c|c c}\hline
system & embedding size & mask & minDCF & EER \\\hline
 Vox1-MFB& 128 & no & 0.721 & 6.49 \\
-Resnet34- & 128 & yes &0.690  & 6.22 \\
FFSVC405 & 512 & no & 0.719 & 6.49\\
 -MFB-fintune & 512 & yes & 0.677 & 6.28 \\\hline
 & 128 & no & 0.738 & 7.16 \\
FFSVC405 & 128 & yes & 0.705 &6.75 \\
-MFB-Resnet34 & 512 & no & 0.729 & 6.98 \\
 & 512 & yes & 0.722 & 6.78 \\\hline
\end{tabular}}
\end{center}
\vspace{-5mm}
\end{table}

\textbf{Vox1-MFB-Resnet34-FFSVC1215-MFB-finetune:} This extractor is similar to Vox1-MFB-Resnet34-FFSVC405-MFB-finetune, with the difference of using FFSVC1215-MFB to finetune.

\textbf{AIshell-MFB-Resnet34-FFSVC1215-MFB-finetune:} This extractor is similar to Vox1-MFB-Resnet34-FFSVC1215-MFB-finetune, with the difference of using AIshell-MFB, 2401 spks, to pre-train the model.

\textbf{FFSVC1215-MFB-Resnet34:} This extractor is similar to FFSVC405-MFB-Resnet34, with the difference of using FFSVC1215-MFB to finetune.

\textbf{FFSVC1215-MFB-Res2net50:} This extractor is similar to FFSVC1215-MFB-Resnet34, with the difference of using Res2net50 embedding extractor, with \textit{width} 16 and \textit{scale} 4.

\textbf{FFSVC1620-MFB-Resnet34:} This extractor is similar to FFSVC405-MFB-Resnet34, with the difference of using FFSVC1215-MFB to finetune.

\textbf{FFSVC1620-MFB-Res2net50:} This extractor is similar to FFSVC1215-MFB-Resnet34, with the difference of using Res2net50 embedding extractor.

The results of this six systems are summarized in Table~\ref{R2}. It could find that the dataset used for pre-training has little effect on the final performance, and finetune is no longer effective when the number of speakers in datasets is comparable to the pre-training set. It is interesting that speed perturbation with speed factors of 0.9 and 1.1 achieves about 10\% relative improvement in EER and 6\% improvement in minDCF. Therefore, we draw the conclusion that speed perturbation is an effective data augmentation method to prevent over-fitting when numbers of speakers is not enough. In addition, Res2net blocks improves the performance by 3\% relatively compared to Resnet blocks. Res2net blocks could get more receptive field to improve the multi-scale feature extraction ability by using \textit{scale} and \textit{width} so that Res2net blocks obtains better performance than Resnet blocks.

\begin{table}[!hbt]
\vspace{-3mm}
\centering
\caption{Results of series of Resnet systems }\label{R2}
\vspace{-5mm}
\begin{center}
\small
\setlength{\tabcolsep}{3mm}{
\begin{tabular}{c| c c}\hline
system & minDCF & EER \\\hline
Vox1-MFB-Resnet34-\\FFSVC1215-MFB-finetune & 0.689 & 6.13 \\\hline
AIshell-MFB-Resnet34-\\FFSVC1215-MFB-finetune &0.695  & 6.02 \\\hline
FFSVC1215-MFB-Resnet34 & 0.648 & 5.62\\\hline
FFSVC1215-MFB-Res2net50 & 0.630 & 5.49 \\\hline
FFSVC1620-MFB-Resnet34 & 0.639 & 5.37 \\\hline
FFSVC1620-MFB-Res2net50 & 0.620 & 5.30 \\\hline
\end{tabular}}
\end{center}
\vspace{-8mm}
\end{table}

\subsection{TDNN systems used Kaldi}
All systems in this series are implemented with Kaldi and MFCC are used as input feature. In addition, all embeddings are transformed to 200 dimension using LDA, and then unit-length normalization and standard PLDA are applied. LDA/PLDA is trained on the finetune dataset. Here, \textit{xvector} represents the common TDNN xvector model.

\textbf{AIshell-MFCC-xvector-FFSVC1620-MFCC-finetune:} According to the Resnet series systems, AIshell-MFCC is used to pretrain xvector model, then FFSVC1620-MFCC is used to finetune. The model is finetuned with an initial learning rate of 0.001 and a final learning rate of 0.0001.

\textbf{AIshell-MFCC-xvector-FFSVC405-MFCC-finetune:} This extractor is similar to AIshell-MFCC-xvector-FFSVC1620 -MFCC-finetune, with the difference of using FFSVC405-MFCC to finetune.

\textbf{AIshell-MFCC-ETDNN-FFSVC1620-MFCC-finetune:} This extractor is similar to AIshell-MFCC-xvector-FFSVC1620 -MFCC-finetune. However, the model to pre-train is ETDNN xvector.

\begin{table*}[!hbtp]
  \caption{Research of backend technique}
  \label{R5}
  \vspace{-5mm}
  \begin{center}
  \centering
  \small
  \setlength{\tabcolsep}{3mm}
{\begin{tabular}{c|c c|c c|c c}\hline
\multirow{2}{*}{System} & \multicolumn{2}{c}{cosine} & \multicolumn{2}{c}{PLDA scoring} & \multicolumn{2}{c}{Asnorm} \\\cline{2-7} 
   & minDCF & EER & minDCF & EER &minDCF & EER \\\hline
 FFSVC1620-MFB-Resnet34  & 0.639 & 5.37 & 0.630 & 5.33  & 0.561 & 5.24  \\\hline
 FFSVC1620-MFB-Res2net50 & 0.620 & 5.30 & 0.621 & 5.22  & 0.547 & 5.16 \\\hline
 AIshell-MFCC-xvector-FFSVC1215-MFCC-finetune & - & - & 0.576 & 5.20 & 0.514 & 4.67 \\\hline
 FFSVC1215-MFB-FTDNN-CEAMS & 0.555 & 5.19 & 0.534 & 4.72 & 0.508 & 4.61 \\\hline
\end{tabular}}{}
\end{center}
\vspace{-8mm}
\end{table*}

\textbf{AIshell-MFCC-xvector-FFSVC1215-MFCC-finetune:} This extractor is similar to AIshell-MFCC-xvector-FFSVC1620 -MFCC-finetune, with the difference of the finetune dataset. 
Only part of FFSVC1215-MFCC is used to finetune the base model. During the finetune stage, the accuracy of the training dataset increases rapidly to 99\%, and the valid dataset is also rapidly maintained at 96\%. So, we attempt to increase difficulty by remove the clean data from close-talking mic in the HIMIA dataset. Meanwhile, considering to channel mismatch between the close-talking enrollment utterance from cellphone and the far-field testing speech from microphone, the recordings from 25cm distance cellphone in The DMASH dataset is not removed. In addition, the impact of the learning rate during finetune stage is explored in our experiments. we attempt to fix the first 6 layers with the learning rate of 0.001-0.0001, and then we untie the fixed layer with the learning rate of 0.0001-0.00005, we named this system AIshell-MFCC-xvector-FFSVC1215-MFCC-finetune-V02. The results of these five systems are shown in Table~\ref{R3}. 

\begin{table}[!hbt]
\vspace{1mm}
\centering
\caption{Results of series of TDNN-Kaldi systems }\label{R3}
\vspace{-5mm}
\begin{center}
\small
\setlength{\tabcolsep}{3mm}{
\begin{tabular}{c| c c}\hline
system & minDCF & EER \\\hline
AIshell-MFCC-xvector-\\FFSVC405-MFCC-finetune & 0.768 & 7.36 \\\hline
AIshell-MFCC-xvector-\\FFSVC1620-MFCC-finetune & 0.612 & 5.82 \\\hline
AIshell-MFCC-ETDNN-\\FFSVC1620-MFCC-finetune &0.671  & 6.21 \\\hline
AIshell-MFCC-xvector-\\FFSVC1215-MFCC-finetune & 0.576 & 5.20\\\hline
AIshell-MFCC-xvector-\\FFSVC1215-MFCC-finetune-V02 & 0.599 & 5.52 \\\hline
\end{tabular}}
\end{center}
\vspace{-2mm}
\end{table}

\subsection{TDNN systems used pyTorch}
All systems in this series are implemented with pyTorch and use MFB as input feature. Particularly, \textit{AMS} represents the AMsoftmax loss is adopted in model training, \textit{CEAMS} represents that the CE-loss branch is added when we train the models using AMsoftmax loss.

\textbf{FFSVC1215-MFB-ETDNN-AMS:} This extractor is similar to FFSVC1215-MFB-Resnet34, with the difference of using ETDNN model and AMsoftmax loss function. Parameters $m$ and $s$ are respectively equal to 0.1 and 30 during the whole training stage. In addition, weight decay uses 1e-3.

\textbf{FFSVC1215-MFB-FTDNN-AMS:} This extractor is similar to FFSVC1215-MFB-ETDNN-AMS, with the difference of using FTDNN model.

\textbf{FFSVC1215-MFB-FTDNN-CEAMS:} This extractor is similar to FFSVC1215-MFB-ETDNN-AMS, with the difference of using CEAMsoftmax loss. During the training stage using AMsoftmax loss, the training process is extremely unstable, and is very prone to overfit, especially when parameters $m$ is greater than 0.1. 
Although the copies which are created by using speed perturbation is considered to belong to diffrent speakers, it's similar to the original signal to some extent. In addition, the deep network is prone to overfit when small-scale training data is available, not to mention similar uttrances. 
To address this problems, CE-loss is added to the network to guide and assist the model training in the early stage. We first add a CE-loss class prediction branch to conventional feedforward architecture in parallel with the AMsoftmax-loss class prediction branch, in practice, the CE-loss class prediction branch is added after the statistics pooling layer of the FTDNN network, and has the same network structure as the AMsoftmax-loss branch. The weight of CE-loss is set to 1.0, 0.5, 0.1 and is switched when the training loss plateaus.

\textbf{FFSVC1620-MFB-FTDNN-CEAMS:} This extractor is similar to FFSVC1215-MFB-ETDNN-AMS, with the difference of using FFSVC1620-MFB.

\begin{table}[!hbt]
\vspace{-3mm}
\centering
\caption{Results of series of TDNN-pyTorch systems }\label{R4}
\vspace{-5mm}
\begin{center}
\small
\setlength{\tabcolsep}{2.4mm}{
\begin{tabular}{c| c c}\hline
system & minDCF & EER \\\hline
FFSVC1215-MFB-ETDNN-AMS & 0.698 & 6.57 \\\hline
FFSVC1215-MFB-FTDNN-AMS & 0.609 & 6.27 \\\hline
FFSVC1215-MFB-FTDNN-CEAMS & 0.555 & 5.21 \\\hline
FFSVC1620-MFB-FTDNN-CEAMS & 0.534 & 5.18 \\\hline
\end{tabular}}
\end{center}
\vspace{-6mm}
\end{table}

The results of the above four models are summarized in Table~\ref{R4}. During the experiment, the model training stability has been greatly improved after adding CE-loss branch, and it achieves about 8\% relative improvement in minDCF and 16\% improvement in EER. 

\subsection{ Backend}
We select four of the best models to do backend scoring. The methods we use includes LDA, PLDA, ASnorm and score fusion. All of the four models use PLDA scoring and ASnorm. 
Firstly, for systems implemented with pytorch, we transform the centered, whitened, and unit-length normalized embeddings by LDA, without dimensionality reduction. Then, one Gaussian PLDA is trained either on FFSVC1215 or FFSVC1620 set. Finally, we use ASnorm to post-process the PLDA verification.
We use the eval set and dev set as each other's cohort set. For each enrollment from 25cm distance cellphone, we randomly select 2700 single far-field microphone arrays from the cohort set to score with it, each testing utterance is similar. Then we select the top 5\% of sorted cohort scores to calculate the normalization. The result of these four systems are shown in Table~\ref{R5}.

\subsection {Submitted system and result}
We submit the fusion results of the above four systems. Considering that task1 and task3 are similar, we use the same models in these two tasks. These four systems are renamed as follows, FFSVC1620-MFB-Resnet34 named system A,FFSVC1620-MFB-Res2net50 named system B, AIshell-MFCC-xvector-FFSVC1215-MFCC-finetune named system C, FFSVC1215-MFB-FTDNN-CEAMS named system D. The fusion results are summarized in Table~\ref{R6}.

\begin{table}[!hbtp]
  \vspace{-2mm}
  \caption{results of submitted systems }
  \label{R6}
  \vspace{-6mm}
  \begin{center}
  \small
  \setlength{\tabcolsep}{1.0mm}
  \scalebox{1.0}
{\begin{tabular}{c| cc | cc | cc}\hline
\multirow{2}{*}{System} & \multicolumn{2}{c}{task1 dev} & \multicolumn{2}{c}{task1 eval} & \multicolumn{2}{c}{task3 eval} \\\cline{2-7} 
   & minDCF & EER & minDCF & EER &minDCF & EER \\\hline
 A+C+D  &  0.442 &  3.88 & 0.549 & 5.08  & 0.526 & 5.61  \\\hline
 A+B+C+D & 0.427 & 3.70  & 0.534 & 5.05  & 0.519 & 5.47 \\\hline
\end{tabular}}{}
\end{center}
\vspace{-8mm}
\end{table}

\section{Conclusions}

In this paper, we detail every aspects of our systems submitted to FFSVC2020 challenge. We describe the usage of data sets, the strategy of data augmentation, various embedding extractors, backend technique. Our data augmentation strategy is greatly helpful when only small-scale training data is available. In addition, AMsoftmax loss will be easier to ultilize after adding CE-loss branch. Res2net block could improve performance by getting multi-scale feature infomation. The above three points might be useful to other researchers. Finally, we don't make full use of the infomation between multi-channel, we will explore how to ultilize these infomations in the future. 

\section{Acknowledgements}
This work is partially supported by the National Natural Science Foundation of China (Nos. 11590772,11590770).

\bibliographystyle{IEEEtran}

\bibliography{mybib}


\end{document}